\documentclass[aps,preprint]{revtex4}%
\usepackage{amsfonts}
\usepackage{amsmath}
\usepackage{amssymb}
\usepackage{graphicx}%
\providecommand{\U}[1]{\protect\rule{.1in}{.1in}}

\begin{document}
\title[ ]{Kratzer's molecular potential in quantum mechanics with a generalized
uncertainty principle}
\author{Djamil Bouaziz}
\email{djamilbouaziz@univ-jijel.dz}
\affiliation{Laboratoire de Physique Th\'{e}orique, D\'{e}partement de Physique,
Universit\'{e} de Jijel, BP 98, Ouled Aissa, 18000 Jijel, Algeria}
\keywords{Minimal length, generalized uncertainty principle, Kratzer potential}
\pacs{PACS number(s) 03.65.Ge, 03.65.Ca, 02.40.Gh, 34.20.Cf}

\begin{abstract}
The Kratzer's potential $V(r)=g_{1}/r^{2}-g_{2}/r$ is studied in quantum
mechanics with a generalized uncertainty principle, which includes a minimal
length $\left(  \Delta X\right)  _{\min}=\hbar\sqrt{5\beta}$. In momentum
representation, the Schr\"{o}dinger equation is a generalized Heun's
differential equation, which reduces to a hypergeometric and to a Heun's
equations in special cases. We explicitly show that the presence of this
finite length regularizes the potential in the range of the coupling constant
$g_{1}$ where the corresponding Hamiltonian is not self-adjoint. In coordinate
space, we perturbatively derive an analytical expression for the bound states
spectrum in the first order of the deformation parameter $\beta$. We
qualitatively discuss the effect of the minimal length on the
vibration-rotation energy levels of diatomic molecules, through the Kratzer
interaction. By comparison with an experimental result of the hydrogen
molecule, an upper bound for the minimal length is found to be of about $0.01$
\AA . We argue that the minimal length would have some physical importance in
studying the spectra of such systems.

\end{abstract}
\volumeyear{2014}
\volumenumber{number}
\issuenumber{number}
\eid{identifier}
\date[Date text]{date}
\received[Received text]{date}

\revised[Revised text]{date}

\accepted[Accepted text]{date}

\published[Published text]{date}

\startpage{1}
\maketitle

\section{Introduction }

The hypothesis of the existence of a minimal length scale \cite{snider,wigner}
is currently a common finding of several studies in quantum gravity
\cite{alden,garay,rov} and string theory \cite{21,amati,konishi}. This
elementary length is supposed to be of the order of the Planck length
($l_{p}=10^{-35}$m), which is a lower bound to all physical length scales,
below which distances can not be resolved \cite{pad,jaekel}. For a review of
the history and the main motivations for the assumption of a fundamental
length scale, see Ref. \cite{hossenfelder}.

The minimal length is introduced in quantum theory by modifying the standard
Heisenberg uncertainty principle to the so-called generalized uncertainty
principle (GUP) \cite{konishi,magiore}. This GUP implies then a non-zero
minimal uncertainty in position, and it involves significant consequences on
the mathematical basis of quantum mechanics. The formalism, based on a
specific form of the GUP, together with the new concepts it implies, has been
extensively discussed by Kempf and co-workers in Refs. \cite{k1,k2,k3,k4}.
Most notably, it has been shown that the GUP can be obtained from a deformed
Heisenberg algebra, depending on two small positive parameters $\beta$ and
$\beta^{\prime}$, which are related to the minimal length (see, Sec. II). For
the sake of completeness, let us mention that diverse forms of the GUP have
been proposed in the literature: there is a GUP which incorporates a minimal
length and a minimal momentum \cite{k3}, a GUP with a Lorentz-covariant
algebra \cite{quesne}, and a GUP including a minimal length and a maximal
momentum \cite{ali,pou}.

In recent years, various topics were studied in connection with the GUP. Among
others, the Schr\"{o}dinger equation for : the harmonic oscillator
\cite{k1,brau,chang}, the hydrogen atom, in one \cite{fit,p} and three
\cite{brau,sandor,stetsko,boua3} dimensions, the singular inverse square
potential \cite{boua1,boua2}, and the gravitational quantum well
\cite{brau2,N}. The minimal length was furthermore introduced in the Dirac
equation, with a constant magnetic field \cite{dirac}, with vector and scalar
linear potentials \cite{cha}, with the hydrogen atom potential \cite{rh}, and
the Dirac oscillator \cite{do}. The effect of the GUP was also studied in the
context of the Klein-Gordon equation in Refs. \cite{KG1,KG2}. Furthermore, the
Casimir effect has been investigated in Ref. \cite{casimir}, the correction to
the black body radiation due to a minimal length has been considered in Ref.
\cite{black-body}, thermostatistic with minimal length has been studied in
Ref. \cite{therm}, Unruh effect with a GUP has been discussed in Ref.
\cite{unr}, and finally, the reconciliation of the Self-Completeness of
gravity and the GUP has been addressed in Ref. \cite{comp}. For a large number
of references in connection with this subject, see, Ref. \cite{hossenfelder}.

It is important to note that although the minimal length is initially
introduced in quantum theory to account for quantum gravitational effects,
which arise at Planck's scales \cite{small}, it has been argued that, in
quantum mechanics, this elementary length may be associated to the size of the
system under study \cite{k2,boua1}; consequently, the formalism that follows
may be useful in the study of complex systems such as quasiparticles, nuclei,
and molecules \cite{k2}. Another important feature of this finite length is
that it provides a natural cutoff for ultraviolet regularization in quantum
mechanics \cite{boua1} and in quantum field theory \cite{k4}.

In this work, we investigate another problem within the formalism of quantum
mechanics with a minimal length, namely the Kratzer's molecular potential
(KP), which has the form $V(r)=g_{1}/r^{2}-g_{2}/r$. This potential is one of
the most important model interaction in quantum physics; it has been first
introduced to describe the vibration-rotation spectra of diatomic molecules
\cite{kratzer,flug}. Nowadays, KP appears in various fields of physics and
chemistry such as molecular physics \cite{haj}, nuclear physics \cite{lom},
Liquid-solid interfaces and thermodynamics \cite{rom}, chemical physics
\cite{hoo}, and quantum chemistry \cite{coo,ber,wal,sec,oye}. From a more
formal viewpoint, KP provides a good example for illustrating diverse methods
used to solve the Schr\"{o}dinger equation, such as the Fourier integral
representation \cite{fi}, the algebraic approach \cite{aa}, the supersymmetry
\cite{coo}, the Nikiforov-Uvarov \cite{ber}, the asymptotic iteration method
\cite{im}, and the method of self-adjoint extensions \cite{self}, which is
used when the coupling constant $g_{1}$ is such as $2\mu g_{1}/\hbar^{2}%
\leq-1/4$ ($\mu$ is the particle mass), where the corresponding Hamiltonian
operator is not self-adjoint \cite{self}.

The aim of this study is twofold, first to examine to what extend the
introduction of a minimal length in the Schr\"{o}dinger equation regularizes
the KP in the aforementioned range of the coupling constant $g_{1}$ where the
potential is known to be singular, and second, to compute the bound states
energy spectrum, and investigate the effect of this fundamental length on the
rovibrational energy levels of diatomic molecules.

The rest of this paper is organized as follows. In Sec. II, we present the
main equations of quantum mechanics with a minimal length, that we need in our
study. Sec. III is devoted to investigate the KP in this formalism, we study
in detail the corresponding Schr\"{o}dinger equation in momentum
representation, then we perturbatively compute the energy spectrum in
coordinate space and apply our result to the vibration-rotation of diatomic
molecules. In the last section, we summarize our results and conclusions.

\section{Heisenberg algebra and representations of the GUP}

As mentioned in Sec. I, diverse topics were studied in quantum mechanics based
on the following \textit{3}-dimensional modified Heisenberg algebra
\cite{k1,k2,chang,boua1,boua2}:%

\begin{align}
\lbrack\widehat{X}_{i},\widehat{P}_{j}]  &  =i\hbar\lbrack(1+\beta\widehat
{P}^{2})\delta_{ij}+\beta^{^{\prime}}\widehat{P}_{i}\widehat{P}_{j}],\text{
\ \ }(\beta,\beta^{^{\prime}})>0.\nonumber\\
\lbrack\widehat{P}_{i},\widehat{P}_{j}]  &  =0,\label{1}\\
\lbrack\widehat{X}_{i},\widehat{X}_{j}]  &  =i\hbar\frac{2\beta-\beta
^{^{\prime}}+\beta(2\beta+\beta^{^{\prime}})\widehat{P}^{2}}{1+\beta
\widehat{P}^{2}}(\widehat{P}_{i}\widehat{X}_{j}-\widehat{P}_{j}\widehat{X}%
_{i}).\nonumber
\end{align}

\bigskip These commutation relations lead to the following generalized
uncertainty principle (GUP) \cite{boua1}:
\begin{equation}
\left(  \Delta X_{i}\right)  \left(  \Delta P_{i}\right)  \geq\frac{\hbar}%
{2}\left(  1+\beta\sum\limits_{j=1}^{N}[\left(  \Delta P_{j}\right)
^{2}+\left\langle \widehat{P}_{j}\right\rangle ^{2}]+\beta^{^{\prime}}[\left(
\Delta P_{i}\right)  ^{2}+\left\langle \widehat{P}_{i}\right\rangle
^{2}]\right)  , \label{2}%
\end{equation}
which implies the existence of a minimal length, given by \cite{k2}
\begin{equation}
\left(  \Delta X_{i}\right)  _{\min}=\hbar\sqrt{3\beta+\beta^{^{\prime}}%
},\text{ \ \ }\forall i. \label{3}%
\end{equation}

One of the more important implications of the GUP is the loss of localization
in coordinate space as effect of the nonzero minimal uncertainty in position
measurements. Consequently, momentum space would be more convenient for
solving quantum mechanical problems.

In the literature, one of the most used representations of the position and
momentum operators satisfying the commutation relations (\ref{1}) is
\cite{k2,chang,boua1}%
\begin{equation}
\widehat{X}_{i}=i\hbar\lbrack\left(  1+\beta p^{2}\right)  \frac{\partial
}{\partial p_{i}}+\beta^{\prime}p_{i}p_{j}\frac{\partial}{\partial p_{j}%
}+\gamma p_{i}],\text{ \ \ \ \ \ }\widehat{P}_{i}=p_{i}, \label{7}%
\end{equation}
where $\gamma$ is a small positive parameter related to $\beta$ and
$\beta^{\prime}$.

The disadvantage of representation (\ref{7}) is that it considerably
complicates the Schr\"{o}dinger equation, and its solution is not often
possible, especially when the potential depends on the position operators in a
not too simple way.

To overcome this difficulty, one can alternatively use a perturbative approach
to include the minimal length in the Schr\"{o}dinger equation. So, coordinate
space would be more appropriate in this case. This approach was first proposed
by Brau to deal with the hydrogen atom problem \cite{brau}. The simplest
representation of the operators $\widehat{X}_{i}$ and $\widehat{P}_{i}$ in
coordinate space is \cite{brau}%
\begin{equation}
\widehat{X}_{i}=\widehat{x}_{i},\text{ \ \ \ }\widehat{P}_{i}=\widehat{p}%
_{i}\left(  1+\beta\widehat{p}^{2}\right)  , \label{brau}%
\end{equation}
where $\widehat{x}_{i}$ and $\widehat{p}_{i}$ satisfy the standard commutation
relations of ordinary quantum mechanics.

Representation (\ref{brau}), which is valid in the case $\beta^{\prime}%
=2\beta$ up to the first order of $\beta$, is mostly more adequate when
applying the perturbation theory to study the deformed Schr\"{o}dinger
equation for a given interaction. As mentioned by Kempf \cite{k2}, this
special case is of particular importance as the commutation relations between
the position operators are not modified at the first order in $\beta$, so that
the space remains commutative. The deformed algebra (\ref{1}) takes then the
form%
\begin{align}
\lbrack\widehat{X}_{i},\widehat{P}_{j}]  &  =i\hbar\lbrack(1+\beta\widehat
{P}^{2})\delta_{ij}+2\beta\widehat{P}_{i}\widehat{P}_{j}],\nonumber\\
\text{ }[\widehat{P}_{i},\widehat{P}_{j}]  &  =0,\text{ \ \ \ \ \ \ }%
[\widehat{X}_{i},\widehat{X}_{j}]=0, \label{4}%
\end{align}
and the minimal length reads in 3-dimensions as
\[
\left(  \Delta X_{i}\right)  _{\min}=\hbar\sqrt{5\beta},\text{ \ \ }\forall
i.
\]
The commutators (\ref{4}) constitute the minimal extension of the Heisenberg
algebra \cite{k2}.

\section{Kratzer potential with a minimal length}

While several potentials were studied in quantum mechanics with the modified
algebra (\ref{1}), the KP is considered here. We discuss in detail its
corresponding deformed Schr\"{o}dinger equation in momentum representation, we
shall be, in particular, interested to the singularity structure of this
equation. Then, we perturbatively compute in coordinate space the bound states
energy spectrum, and apply our result to the vibration-rotation motion of
diatomic molecules.

\subsection{Schr\"{o}dinger equation in momentum representation}

The Kratzer potential that we consider in this work has the form
\cite{kratzer,flug}%
\begin{equation}
V(r)=D_{e}r_{e}(r_{e}/r^{2}-2/r), \label{2.1}%
\end{equation}
where $D_{e}$ is the dissociation energy and $r_{e}$ is the equilibrium
internuclear distance of a given diatomic molecule.

First, for convenience and for a later comparison with the hydrogen atom and
the inverse square interactions, we write the KP in the general form%
\begin{equation}
V(r)=g_{1}/r^{2}-g_{2}/r. \label{kr}%
\end{equation}

We proceed our study by writing the Schr\"{o}dinger equation for the energy
function (\ref{kr}) in momentum representation as follows:%

\begin{equation}
\left(  \widehat{R}^{2}(\frac{1}{2\mu}p^{2}-E)-g_{2}\widehat{R}+g_{1}\right)
\psi(p)=0, \label{sch}%
\end{equation}
where $\widehat{R}^{2}\equiv\widehat{R}\times\widehat{R}=%
{\textstyle\sum\limits_{i=1}^{3}}
\widehat{X}_{i}\widehat{X}_{i}$, and $\mu$ is the particle mass (the reduced
mass of the two atoms, in the study of diatomic molecules).

In general, the deformed Schr\"{o}dinger equation (\ref{sch}) can not be
established with the momentum representation (\ref{7}) of the operators
$\widehat{X}_{i}$. The difficulty lies in the definition of the operator
$\widehat{R}$ because the factorization of $\ \widehat{R}^{2}$ is not obvious.
However, it has been shown in Ref. \cite{boua3} that, in the case
$\beta^{\prime}$ $=2\beta$, this operator can be factorized in the first order
of $\beta$ for the $s$-waves, and hence the Schr\"{o}dinger equation can be
written in this special case.

In effect, by restricting ourselves to the $\ell=0$ wave function and by using
the momentum representation (\ref{7}), with $\gamma=0$, we obtain the
following expression for the distance squared operator \cite{boua3}:%
\begin{equation}
\widehat{R}^{2}=\left(  i\hbar\right)  ^{2}\left\{  \left[  1+\left(
\beta+\beta^{\prime}\right)  p^{2}\right]  ^{2}\frac{d^{2}}{dp^{2}}+\frac
{2}{p}\left[  1+\left(  \beta+\beta^{\prime}\right)  p^{2}\right]  \left[
1+\left(  2\beta+\beta^{\prime}\right)  p^{2}\right]  \frac{d}{dp}\right\}  .
\label{R2}%
\end{equation}

Since $\beta$ and $\beta^{\prime}$ are supposed to be small parameters, the
distance squared operator can be expressed, in the case $\beta^{\prime}$
$=2\beta$, as%
\begin{equation}
\widehat{R}^{2}\mathbf{=}\left(  i\hbar\right)  ^{2}\left\{  (1+6\beta
p^{2})\frac{d^{2}}{dp^{2}}+\frac{2}{p}\allowbreak(1+7\beta p^{2})\frac{d}%
{dp}\right\}  +O\left(  \beta^{2}\right)  . \label{R1}%
\end{equation}
Now, with Eq. (\ref{R1}), the square root operator $\widehat{R}$ can be
defined. Thus, it is easy to check that $\widehat{R}^{2}$ can be written as
$\widehat{R}\times\widehat{R}$\textbf{,} where%
\begin{equation}
\widehat{R}\mathbf{=}i\hbar\left[  \left(  1+3\beta p^{2}\right)  \frac{d}%
{dp}+\frac{1}{p}\allowbreak\allowbreak\left(  1+\beta p^{2}\right)  \right]
+O\left(  \beta^{2}\right)  . \label{R}%
\end{equation}

By using formulas (\ref{R1}) and (\ref{R}) in Eq. (\ref{sch}), we obtain the
following differential equation:%

\begin{align}
&  \left(  p^{2}+k^{2}\right)  (1+6\beta p^{2})\frac{d^{2}\psi}{dp^{2}%
}+\left\{  2\allowbreak(3+19\beta p^{2})p+\frac{2k^{2}}{p}\allowbreak(1+7\beta
p^{2})+2i\sigma_{2}\left(  1+3\beta p^{2}\right)  \right\}  \frac{d\psi}%
{dp}\nonumber\\
&  +\left\{  2(3+20\beta p^{2})+\frac{2i\sigma_{2}}{p}\allowbreak
\allowbreak\left(  1+\beta p^{2}\right)  -\sigma_{1}\right\}  \psi=0,
\label{d}%
\end{align}
where we have used the notations%
\[
\text{ }k^{2}=-2\mu E,\text{ \ \ \ \ }\sigma_{1}=\frac{2\mu g_{1}}{\hbar^{2}%
},\text{ \ \ \ \ }\sigma_{2}\text{\ }=\frac{\mu g_{2}}{\hbar}.
\]

The singularity of Eq. (\ref{d}) in $p=0$ can be removed by performing the
following change of function:%
\[
\psi(p)=\frac{1}{p}\varphi(p).
\]
This leads to the differential equation%
\begin{align}
&  (p^{2}+k^{2})(1+6\beta p^{2})\dfrac{d^{2}\varphi}{dp^{2}}+\left\{  2\beta
p(p^{2}+k^{2})+4p(1+6\beta p^{2})+2i\sigma_{2}(1+3\beta p^{2})\right\}
\dfrac{d\varphi}{dp}\nonumber\\
&  +\left\{  4(1+7\beta p^{2})-2\beta(p^{2}+k^{2})-2(1+6\beta p^{2}%
)-4i\beta\sigma_{2}p-\sigma_{1}\right\}  \varphi=0. \label{c}%
\end{align}

Equation (\ref{c}) is the Schr\"{o}dinger equation for the KP in momentum
space with a minimal length $\left(  \Delta X_{i}\right)  _{\min}=\hbar
\sqrt{5\beta}$. Before studying this equation, it is important to consider the
undeformed case, i.e., the limit $\beta=0.$

\subsubsection{Undeformed case}

In the special case $\beta=0$, Eq. (\ref{c}) reduces to the following
equation:%
\begin{equation}
(p^{2}+k^{2})\dfrac{d^{2}\varphi}{dp^{2}}+(4p+2i\sigma_{2})\dfrac{d\varphi
}{dp}+\left(  2-\sigma_{1}\right)  \varphi=0. \label{ud}%
\end{equation}
By using the new variable%
\[
x=1/2+ip/2k,
\]
equation (\ref{ud}) can be transformed to a hypergeometric differential
equation of the form \cite{abramo}%
\begin{equation}
x(1-x)\varphi^{\prime\prime}+\left[  c-(a+b+1)x\right]  \varphi^{\prime
}-ab\varphi=0, \label{h}%
\end{equation}
with the parameters%
\begin{align*}
a  &  =3/2+\nu,\text{ \ \ \ \ \ \ }\nu=\sqrt{1/4+\sigma_{1}}\\
b  &  =3/2-\nu,\text{ \ \ \ \ \ \ }c=2+\sigma_{2}/k.\text{ \ \ \ \ \ \ }%
\end{align*}

In the vicinity of infinity, the two linearly independent solutions of Eq.
(\ref{h}) are \cite{abramo}%
\begin{align*}
\varphi_{1}\left(  y\right)   &  =x^{-a}F\left(  a,a-c+1,a-b+1;1/x\right)  ,\\
\varphi_{2}\left(  y\right)   &  =x^{-b}F\left(  b,b-c+1,b-a+1;1/x\right)  ,
\end{align*}
where $F\ $stands for the hypergeometric function.

The two solutions of the Schr\"{o}dinger equation behave then, in the limit
($p\gg1$), as%
\begin{align}
\psi_{1}\left(  p\right)   &  \rightarrow p^{-5/2-\nu},\text{ \ \ \ }%
\label{s1}\\
\text{\ }\psi_{2}\left(  p\right)   &  \rightarrow p^{-5/2+\nu}. \label{s2}%
\end{align}

In the case where $\nu$ is real ($\sigma_{1}>-1/4$), the function $\psi_{2}$
falls off more slowly than $\psi_{1}$, and hence the behavior of the physical
solution is that of $\psi_{1}$. This may be understood by the fact that the
expectation value of $\widehat{R}^{2}$ diverges with the function $\psi_{2}$,
as in the case of the relativistic Coulomb problem \cite{bs}.

Then, the solution to the ordinary Schr\"{o}dinger equation for the KP in
momentum space reads%
\begin{equation}
\psi_{1}\left(  p\right)  =\frac{N}{p}\left(  1+ip/k\right)  ^{-\frac{3}%
{2}-\nu}F\left(  \frac{3}{2}+\nu,\frac{1}{2}-\frac{\sigma_{2}}{k}+\nu
,1+2\nu;\frac{2}{1+ip/k}\right)  , \label{5}%
\end{equation}
where $N$ is a normalization constant.

The discrete energy spectrum of the KP can be obtained by requiring%
\begin{equation}
1/2-\sigma_{2}/k+\nu=-n,\text{ \ \ \ \ \ \ \ \ }n=0,1,2..., \label{s}%
\end{equation}
so that the wave function (\ref{5}) is square integrable for the whole
interval of $p$. In this case, the hypergeometric series reduces to a
polynomial \cite{abramo}.

Equation (\ref{s}) is the quantization condition of the energy, it gives the
well-known discrete energy spectrum of the KP (see Eq. (\ref{es}), below).

However, when $\sigma_{1}\leq-1/4$, the parameter $\nu$ becomes imaginary and
thus the spectral condition (\ref{s}) breaks down. Moreover, the asymptotic
behaviors of the two solutions, given by Eqs. (\ref{s1}) and (\ref{s2}),
become identical, and both of them are physical. Consequently, the general
solution is a linear combination of $\psi_{1}$ and $\psi_{2}$, and thus the
wave function and other related observables will now depend on an arbitrary
parameter (phase), which is a common feature of singular potentials
\cite{case}.

We mention that the singularity of the KP is due to the inverse square
interaction term $g_{1}/r^{2}$, which is known to be singular in the strong
coupling regime ($\sigma_{1}=2\mu g_{1}/\hbar^{2}\leq-1/4$) \cite{case}. In
this range of the coupling, the Hamiltonian operator corresponding to KP is
not self-adjoint, one must then define the self-adjoint extensions of the
Hamiltonian, for a recent review, see Ref. \cite{self}.

\subsubsection{Deformed case}

We return now to Eq. (\ref{c}). We first discuss the effect of the minimal
length on the singularity structure of this equation. To this end, let us
write Eq. (\ref{c}) in the limit $p\gg1$ as follows:%
\begin{equation}
3p^{2}\dfrac{d^{2}\varphi}{dp^{2}}+13p\dfrac{d\varphi}{dp}+7\varphi=0,
\label{a}%
\end{equation}
for which the solutions are $\varphi_{1}\sim p^{-7/3}\ $and $\varphi_{2}%
\sim\ p^{-1}$.

The solutions of the deformed Schr\"{o}dinger equation (\ref{d}) behave then
at infinity as%
\[
\psi_{1}\sim p^{-10/3},\ \ \ \ \psi_{2}\sim\ p^{-2}.
\]
These behaviors are completely different from that of the undeformed case,
given by Eqs. (\ref{s1}) and (\ref{s2}). Now, the second solution $\psi_{2}$
falls off more slowly than $\psi_{1}$ regardless the value of the
\textquotedblright coupling constant\textquotedblright\ $\sigma_{1}$. It
follows that the physical solution has manifestly the asymptotic behavior of
$\psi_{1}$; one can always reject $\psi_{2}$ even for $\sigma_{1}\leq-1/4$, so
that the wave function will not depend on an arbitrary phase, unlike those of
singular potentials \cite{case}. Consequently, in the presence of a minimal
length, there is no difference between the two ranges of the coupling $g_{1}$
of the potential: $2\mu g_{1}/\hbar^{2}\leq-1/4$ and $2\mu g_{1}/\hbar
^{2}>-1/4$. This might be viewed as an indication of the regularization of the
KP by the presence of this elementary length. It is important to note that the
rejection of the asymptotic behavior of $\psi_{2}$ can be, convincingly,
reached by requiring that the physical eigenfunctions of the Hamiltonian must
behave at large momenta as $p^{2}\psi(p)_{p\rightarrow\infty}=0$. This
boundary condition emerges naturally from the integral equation corresponding
to the differential equation (\ref{d}); the demonstration is analogous to that
given in detail in Ref. \cite{boua1} for the inverse square potential.

We now discuss the class to which Eq. (\ref{c}) belongs and its associated
properties. After carefully examining this equation, we have shown that it is
a Fuchsian differential equation with five regular singular points. In fact,
by making the change of variable%
\[
z=\frac{1}{2}(1-i\sqrt{6\beta}p),
\]
Eq. (\ref{c}) can be transformed to the following generalized Heun's equation
\cite{hon,wang}:%
\begin{equation}
\dfrac{d^{2}\varphi}{dz^{2}}+\left(  \dfrac{c}{z}+\dfrac{d}{(z-1)}+\dfrac
{e}{(z-z_{1})}+\dfrac{f}{(z-z_{2})}\right)  \dfrac{d\varphi}{dz}+\left(
\dfrac{abz^{2}+\rho_{1}z+\rho_{2}}{z\left(  z-1\right)  (z-z_{1})(z-z_{2}%
)}\right)  \varphi=0. \label{14}%
\end{equation}
The parameters of Eq. (\ref{14}) are given by%
\begin{align*}
a  &  =1,\text{ \ \ \ }b=\frac{7}{3},\text{ \ \ \ \ }\rho_{1}=-\frac{7}%
{3}-\frac{\sigma_{1}}{3}\sqrt{6\beta},\text{ \ \ \ }\rho_{2}=\frac{\beta}%
{2}k^{2}+\frac{\sigma_{1}}{6}\sqrt{6\beta}+\allowbreak\frac{1}{12}%
+\frac{\sigma_{2}}{4},\\
c  &  =\frac{1}{6}+\frac{\sigma_{1}}{2}\frac{\sqrt{6\beta}}{1-6\beta k^{2}%
},\text{ \ \ \ \ }d=\frac{1}{6}-\frac{\sigma_{1}}{2}\frac{\sqrt{6\beta}%
}{1-6\beta k^{2}},\text{ \ \ \ \ \ \ \ }e=2+\frac{\sigma_{1}}{k}\frac{1-3\beta
k^{2}}{1-6\beta k^{2}},\\
f  &  =2-\frac{\sigma_{1}}{k}\frac{1-3\beta k^{2}}{1-6\beta k^{2}},\text{
\ \ \ \ \ \ \ \ \ \ \ }z_{1}=\frac{1}{2}+\frac{k}{2}\sqrt{6\beta},\text{
\ \ \ \ \ \ \ \ \ \ \ \ \ }z_{2}=\frac{1}{2}-\frac{k}{2}\sqrt{6\beta},
\end{align*}
which are linked by the Fuchsian condition%
\[
a+b+1=c+d+e+f.
\]

Equation (\ref{14}) belongs to the class of Fuchsian equations: it is a linear
homogeneous second-order differential equation with five singular points
located at $z=0,$ $1,$ $z_{1},$ $z_{2},$ $\infty$, all regular. So, it admits
power series solutions in the vicinity of each singular point \cite{wang}.
However, to the best of our knowledge, the analytic solutions to the
generalized Heun's equation (\ref{14}) are not known in the literature, i.e.,
the recurrence relation that determines the coefficients of the series was not
established for equations of type (\ref{14}). It follows that the formulation
of a physical problem with this kind of equations is interesting in its own
right. This might motivate profound studies on such type of equations.

To end this section, it is important to mention that in the special case
$\sigma_{2}$\ $=0$, where the KP reduces to the inverse square potential, Eq.
(\ref{d}) can be transformed to a Heun's differential equation of the form
\cite{ronv}%
\begin{equation}
\frac{d^{2}\phi}{d\xi^{2}}+\left(  \frac{c}{\xi}+\frac{d}{\xi-1}+\frac{e}%
{\xi-\xi_{0}}\right)  \frac{d\phi}{d\xi}+\left(  \frac{ab\xi+q}{\xi\left(
\xi-1\right)  \left(  \xi-\xi_{0}\right)  }\right)  \phi=0, \label{11}%
\end{equation}
by using the change of variable $\xi=\frac{6\beta p^{2}}{1+6\beta p^{2}}$, and
the transformation $\psi\left(  \xi\right)  =(1-\xi)\phi\left(  \xi\right)  $,
with the parameters%
\begin{equation}
a=\frac{11}{6},\text{ \ }b=1,\text{ \ }c=\frac{3}{2},\text{ \ }d=\frac{1}%
{3},\text{ \ }e=2,\text{ \ }q=-\frac{3}{2}+\frac{\sigma_{1}/4}{1+12\mu\beta
E},\text{ \ }\xi_{0}=\frac{12\mu\beta E}{1+12\mu\beta E}, \label{pc}%
\end{equation}
where now the Fuchsian condition is $a+b+1=c+d+e.$

The inverse square potential has been already studied in details in Ref.
\cite{boua1} with the exact expression of the operator $\widehat{R}^{2}$,
which depends on two deformation parameters $\beta$ and $\beta^{\prime}$.
Whereas, Eq. (\ref{11}) and their parameters (\ref{pc}) have been obtained by
using the approximate expressions of $\widehat{R}^{2}$ and $\widehat{R}$ of
Ref. \cite{boua3}. However, one can from Eq. (\ref{11}) reach all the results
and conclusions of Ref. \cite{boua1}, which proves the validity of our approximation.

It turns out that Eq. (\ref{c}) possesses interesting features, which would be
important from a mathematical viewpoint. In particular, the reduction of Eq.
(\ref{c}) to a hypergeometric equation in the case $\beta=0$ and to a Heun's
equation when $\sigma_{2}$\ $=0$ may be useful especially when studying the
process of coalescence of the singular points in such generalized Heun's equations.

To complete our study, it is important to investigate the effect of the
minimal length on the energy spectrum of the KP. To this purpose, we will
consider the Schr\"{o}dinger equation in coordinate representation.

\subsection{ Schr\"{o}dinger equation in coordinate space: Energy spectrum}

Let us\ now write the Schr\"{o}dinger equation for the KP (\ref{2.1}) in
coordinate space by using the representation (\ref{brau}) as follow%

\begin{equation}
\left(  \frac{\widehat{p}^{2}}{2\mu}+V(r)+\frac{\beta}{\mu}\widehat{P}%
^{4}\right)  \psi(\overset{\rightarrow}{r})=E\psi(\overset{\rightarrow}{r}),
\label{6}%
\end{equation}
where terms of order $\beta^{2}$ have been neglected.

In the ordinary case ($\beta=0$), Eq. (\ref{6}) with the KP allows its exact
solution for arbitrary quantum number $\ell$. This is an advantage compared to
other molecular interactions such the well-known Morse potential
\cite{flug,morse}, for which the Schr\"{o}dinger equation has exact solution
only in the case $\ell=0$. In coordinate space, the solution of the ordinary
Schr\"{o}dinger equation with the KP can be found in the standard textbooks of
quantum mechanics, see, for instance, \cite{flug,landau}. The bound states
energy eigenvalues and the corresponding normalized eigenfunctions are given
by \cite{flug,landau}:%
\begin{equation}
E_{n\ell}^{0}=-\frac{\gamma^{2}D_{e}}{\left(  \lambda+n\right)  ^{2}},\text{
\ \ \ \ }n=0,1,2,...,\text{ \ \ \ }\ell=0,1,2,..., \label{es}%
\end{equation}%
\begin{equation}
\psi_{n\ell m}^{0}(r,\theta,\varphi)=NY_{\ell}^{m}(\theta,\varphi)\left(
r/r_{e}\right)  ^{\lambda-1}e^{-\alpha r/r_{e}}{}_{1}F_{1}(-n,2\lambda;2\alpha
r/r_{e}), \label{wf}%
\end{equation}
where $n$ and $\ell$ are, respectively, the radial (vibrational) and orbital
(rotational) quantum numbers, $Y_{\ell}^{m}(\theta,\varphi)$ are the
orthonormalized spherical harmonics, ${}_{1}F_{1}(\delta,\eta;z)$ is a
confluent hypergeometric function; we have used the following notations:%
\begin{align*}
\text{\ }\gamma &  =\frac{r_{e}}{\hbar}\sqrt{2\mu D_{e}}=\frac{2D_{e}}%
{\hbar\omega},\text{ \ }\alpha=\frac{\gamma^{2}}{\left(  \lambda+n\right)
},\text{ \ }\\
\text{\ }\lambda &  =1/2+\sqrt{(\ell+1/2)^{2}+\gamma^{2}},\text{ \ \ }%
N=\frac{r_{e}^{-\frac{3}{2}}(2\alpha)^{\lambda+\frac{1}{2}}}{\Gamma(2\lambda
)}\sqrt{\frac{\Gamma(2\lambda+n)}{2n!(\lambda+n)}},
\end{align*}
\ with $\omega\ $is the classical frequency for small harmonic vibrations.

In the undeformed case ($\beta\neq0$), the solution to Eq. (\ref{6}) is not
obvious. However, we can consider, in Eq. (\ref{6}), the term ($\frac{\beta
}{\mu}\widehat{p}^{4}$) as a perturbation to the ordinary Schr\"{o}dinger
equation. Therefore, the use of the perturbation theory allows for the
computation of the corrections to the energy levels in the first order of the
deformation parameter $\beta$.

The energy eigenvalues can then be written as%
\[
E_{n\ell}=E_{n\ell}^{0}+\Delta E_{n\ell},
\]
where $E_{n\ell}^{0}$ are the unperturbed levels given by Eq. (\ref{es}).

In the first order of $\beta$, the correction $\Delta E_{n\ell}$ is then%
\[
\Delta E_{n\ell}=\frac{\beta}{\mu}\langle\psi_{n\ell}^{0}\left\vert
p^{4}\right\vert \psi_{n^{\prime}\ell^{\prime}}^{0}\rangle\equiv\frac{\beta
}{\mu}\langle n\ell m\left\vert p^{4}\right\vert n^{\prime}\ell^{\prime
}m^{\prime}\rangle.
\]
It has been shown in Ref. \cite{brau} that, for central interactions, $\Delta
E_{n\ell}$ can be expressed as follows:%
\begin{equation}
\Delta E_{n\ell}=4\beta\mu\left[  \left(  E_{n\ell}^{0}\right)  ^{2}%
-2E_{n\ell}^{0}\langle n\ell m\left\vert V(r)\right\vert n\ell m\rangle
+\langle n\ell m\left\vert V^{2}(r)\right\vert n\ell m\rangle\right]  .
\label{8}%
\end{equation}

From formula (\ref{8}) with the KP (\ref{kr}) we get%
\begin{align}
\Delta E_{n\ell}  &  =4\beta\mu\left(  \left(  E_{n\ell}^{0}\right)
^{2}+2g_{2}E_{n\ell}^{0}\langle n\ell m\left\vert \frac{1}{r}\right\vert n\ell
m\rangle+(g_{2}^{2}-2g_{1}E_{n\ell}^{0})\langle n\ell m\left\vert \frac
{1}{r^{2}}\right\vert n\ell m\rangle\right. \nonumber\\
&  \left.  -2g_{1}g_{2}\langle n\ell m\left\vert \frac{1}{r^{3}}\right\vert
n\ell m\rangle+g_{1}^{2}\langle n\ell m\left\vert \frac{1}{r^{4}}\right\vert
n\ell m\rangle\right)  . \label{co}%
\end{align}

Thus, to compute the minimal length correction, one has to evaluate the matrix
elements
\[
\langle\frac{1}{r^{p}}\rangle=N^{2}(\frac{2g_{2}}{g_{1}})^{3-p}\int
_{0}^{\infty}y^{2\lambda-p}e^{-2\alpha y}\left(  {}_{1}F_{1}(-n,2\lambda
;2\alpha y)\right)  ^{2}dy,\text{ \ \ \ }p=1,\text{ }2,\text{ }3,\text{ }4,
\]
where we have introduced the dimensionless variable $y=\frac{g_{1}}{2g_{2}}r$. \ 

The computation of the above integrals leads to the following results:
\begin{align*}
\langle\frac{1}{r}\rangle &  =\frac{\mu g_{2}}{\hbar^{2}}\frac{1}%
{(\lambda+n)^{2}},\text{\ \ \ \ \ \ \ \ \ \ \ \ \ \ \ \ \ \ }\langle\frac
{1}{r^{3}}\rangle=(\frac{\mu g_{2}}{\hbar^{2}})^{3}\frac{1}{\lambda
(\lambda-\frac{1}{2})(\lambda-1)(\lambda+n)^{3}}\\
\langle\frac{1}{r^{2}}\rangle &  =(\frac{\mu g_{2}}{\hbar^{2}})^{2}\frac
{1}{(\lambda-\frac{1}{2})(\lambda+n)^{3}},\text{ \ }\langle\frac{1}{r^{4}%
}\rangle=(\frac{\mu g_{2}}{\hbar^{2}})^{4}\frac{1+\frac{3n}{\lambda}\left(
1+\frac{n-1}{2\lambda+1}\right)  }{(\lambda-\frac{1}{2})(\lambda
-1)(\lambda-\frac{3}{2})(\lambda+n)^{5}}.
\end{align*}

By inserting the values of these matrix elements and the expression of
$E_{n,\ell}^{0}$, given by Eq. (\ref{es}), into Eq. (\ref{co}), we get%
\begin{align}
\Delta E_{n\ell}  &  =4\beta\mu^{3}\left(  \frac{g_{2}}{\hbar\left(
\lambda+n\right)  }\right)  ^{4}\left\{  -\frac{3}{4}+\frac{\lambda+n}%
{\lambda-\frac{1}{2}}\left(  1+\frac{\mu g_{1}}{\hbar^{2}}\left(  \frac
{1}{\left(  \lambda+n\right)  ^{2}}-\frac{2}{\lambda(\lambda-1)}\right)
\right)  \right. \nonumber\\
&  \left.  +\left(  \frac{\mu g_{1}}{\hbar^{2}}\right)  ^{2}\frac{1}%
{(\lambda-\frac{1}{2})(\lambda-1)(\lambda-\frac{3}{2})(\lambda+n)}\left(
1+\frac{3n(2\lambda+n)}{\lambda(2\lambda+1)}\right)  \right\}  . \label{cof}%
\end{align}

In the limit $g_{1}=0$, where KP reduces to the Coulomb potential, and hence%
\begin{align*}
\gamma &  =\frac{1}{\hbar}\sqrt{2\mu g_{1}}=0,\text{ }\alpha=0,\text{ }%
\lambda=1/2+\sqrt{(\ell+1/2)^{2}+\gamma^{2}}=\ell+1\\
E_{n\ell}^{0}(g_{1}  &  =0)=-\frac{\mu g_{2}^{2}}{2\hbar^{2}\left(
\ell+1+n\right)  ^{2}}=-\frac{\mu g_{2}^{2}}{2\hbar^{2}n_{p}^{2}}\text{,
\ \ \ \ \ \ \ \ }n_{p}=1,2,...
\end{align*}
where, $n_{p}\ $is the principal quantum number. The correction (\ref{cof})
simplifies then to%
\[
\Delta E_{n\ell}=\frac{4\beta\mu^{3}g_{2}^{4}}{\hbar^{4}n_{p}^{4}}\left(
-3/4+\frac{n_{p}}{\ell+1/2}\right)  ,
\]
which is exactly the result obtained in Ref. \cite{brau} for the hydrogen atom problem.

Finally, the complete energy spectrum of the KP in the presence of a minimal
length can be written in terms of Kratzer's parameters $r_{e}\ $and $D_{e}$
($g_{2}=2D_{e}r_{e},$ \ $g_{1}=D_{e}r_{e}^{2}$) as follows:%
\begin{align}
E_{n\ell}  &  =-\frac{\gamma^{2}D_{e}}{\left(  \lambda+n\right)  ^{2}}%
+\beta\mu D_{e}^{2}\left(  \frac{2\gamma}{\lambda+n}\right)  ^{4}\left\{
-\frac{3}{4}+\frac{\lambda+n}{\lambda-\frac{1}{2}}\left(  1+\frac{\gamma^{2}%
}{2}\left(  \frac{1}{\left(  \lambda+n\right)  ^{2}}-\frac{2}{\lambda
(\lambda-1)}\right)  \right)  \right. \nonumber\\
&  \left.  +\frac{\gamma^{4}}{4}\frac{1}{(\lambda-\frac{1}{2})(\lambda
-1)(\lambda-\frac{3}{2})(\lambda+n)}\left(  1+\frac{3n(2\lambda+n)}%
{\lambda(2\lambda+1)}\right)  \right\}  . \label{sss}%
\end{align}

Formula (\ref{sss}) shows the effect of this deformed algebra on the energy
levels of KP. It can be furthermore used to study several features of diatomic
molecules. In particular, it allows us to investigate the effect of the
minimal length on the rovibrational energy levels of diatomic molecules.

\begin{quote}
An application:Vibration-rotation of diatomic molecules
\end{quote}

As outlined in Sec. I, the KP is one of the most important molecular
interactions; it has long been used to describe the vibration-rotation energy
spectrum of diatomic molecules \cite{kratzer,flug}. The importance of this
potential lies in that its ordinary Schr\"{o}dinger equation admits an exact
solution for arbitrary rotational quantum number $\ell$. This is an advantage
compared to other molecular interactions such as the well-known Morse
potential \cite{flug}. In addition, the energy spectrum of the KP is, as we
will see, similar to the well-known spectroscopic formula \cite{spec}.

Then Eq. (\ref{sss}) can be used to qualitatively investigate the effect of
the minimal length on different parts of the rovibrational energy levels of a
given diatomic molecule with the KP interaction.

For this purpose, following Ref. \cite{boua4}, we use the fact that the
dimensionless parameter $\gamma$, in Eq. (\ref{sss}), is so large for most
molecules ($\gamma\gg1$) \cite{flug}, we may then expand $E_{n\ell}$, given by
Eq. (\ref{sss}), into powers of $1/\gamma$. This leads to the following
expression:%
\begin{align}
E_{n\ell}  &  =D_{e}\left(  -1+2(n+\frac{1}{2})\frac{1}{\gamma}+(\ell+\frac
{1}{2})^{2}\frac{1}{\gamma^{2}}\allowbreak-3(n+\frac{1}{2})^{2}\frac{1}%
{\gamma^{2}}+4(n+\frac{1}{2})^{3}\frac{1}{\gamma^{3}}-3(n+\frac{1}{2}%
)(\ell+\frac{1}{2})^{2}\frac{1}{\gamma^{3}}\right) \nonumber\\
&  +\beta\mu D_{e}^{2}\left(  6\left\{  (n+\frac{1}{2})^{2}+\frac{1}%
{4}\right\}  \frac{1}{\gamma^{2}}\allowbreak+2(n+\frac{1}{2})\left\{
-\frac{1}{4}+4(\ell+\frac{1}{2})^{2}-15(n+\frac{1}{2})^{2}\right\}  \frac
{1}{\gamma^{3}}\right)  +... \label{r}%
\end{align}
This formula shows the different parts of the rovibrational energy spectrum,
and the corrections that the minimal length introduces in each part. The
undeformed spectrum ($\beta=0$) contains three kind of energy: the energy of
harmonic vibrations, that is the second term in Eq. (\ref{r}), the third term
is the rotational energy levels, the fourth and the fifth terms describe the
anharmonicity of vibrations, and the sixth term represents the energy due to
the coupling vibration-rotation. The last remaining terms are corrections due
to the presence of a minimal length.

As we see, the leading correction induced by the minimal length is
proportional to $1/\gamma^{2}$, and depends only on the vibrational quantum
number $n$. So, it affects the anharmonicity of vibrations. The next leading
contribution brought by this elementary length depends on $1/\gamma^{3}$, and
carries three corrections: the first affects the energy of harmonic
vibrations, which is proportional to $1/\gamma$, so its effect would not be
significant. The second term modifies the energy corresponding to the
vibration-rotation coupling , which similarly depends on $1/\gamma^{3}$ in the
ordinary case. The last term proportional to $1/\gamma^{3}$, describes clearly
a second correction to the anharmonicity of vibration; it is then less
important compared to the previous correction proportional to $1/\gamma^{2}$.

Moreover, formula (\ref{r}) might be viewed as an energy spectrum of a
three-parameter potential, i.e., $D_{e},$ $r_{e},$ and $\beta$. In this sense,
the procedure of fitting experimental data used in the case of three-parameter
potentials, such as Morse function, can be followed for adjusting the
parameters of the "deformed KP" with the spectroscopic data of diatomic
molecules. This viewpoint would be important because it can give not only
values of $\beta$ for any molecule, but also the Kratzer model can be improved
with this third parameter. In addition, the introduction of this deformation
parameter in the currently most successful models might help in the search for
an universal potential function of diatomic molecules, which is of fundamental
importance in several areas of physics and chemistry \cite{hu}.

To end this study, it is important to roughly estimate an upper bound for the
minimal length by using formula (\ref{sss}). To this end, we consider the
vibrational ground-state energy $E_{00}$ of the hydrogen molecule (H$_{2}$),
which is one of the more remarkably successful predictions of Kratzer
potential. So, the comparison of Eq. (\ref{sss}) to accurate experimental
results allows to give a value of $\beta$, by assuming that the effect of the
minimal length is included in the gap between the theoretical and the
experimental values of $E_{00}$.

Indeed, the experimental vibrational zero point energy (ZPE$\equiv
G=E_{00}+D_{e}$) for the molecule H$_{2}$ in the state $^{1}\Sigma_{g}^{+}$
has the value \cite{ZPE}%
\[
G^{\text{exp}}=2179.3(1)\text{ cm}^{-1}.
\]
On the other hand, the theoretical value predicted by Kratzer's potential can
be obtained from Eq. (\ref{es}), by using the molecular constants \cite{mc}:
$D_{e}=78844.9005$ cm$^{-1}$, $r_{e}=0.73652$ \AA , and $\mu=0.503\,9$ uma. We
then get the result%
\[
G^{\beta=0}=E_{00}^{0}+D_{e}=2174.\,\allowbreak9\text{ cm}^{-1}.
\]
The theoretical value is not contained in the error bars. therefore, the
discrepancy between $G^{\text{exp}}$ and $G^{\beta=0}$ will be entirely
attributed to the minimal length correction ($\Delta E_{00}$), given by Eq.
(\ref{cof}). We then have the natural constraint%
\begin{equation}
\Delta E_{00}<G^{\text{exp}}-G^{\beta=0}=4.4\text{ cm}^{-1}\approx
5.\,\allowbreak455\,4\times10^{-4}\text{ eV}, \label{di}%
\end{equation}
which leads to the following upper bound of this fundamental length:
\begin{equation}
\left(  \Delta X\right)  _{\min}=\hbar\sqrt{5\beta}<1.\,\allowbreak
0186\times10^{-2}\text{ \AA }\simeq0.01\text{ \AA .} \label{ub}%
\end{equation}

This upper bound is, of course, not a stringent limit, because the Kratzer's
potential is far to be the real internuclear potential energy function in the
molecule H$_{2}$. In addition, many effects, such as relativistic and QED
corrections, are also important and must be included in the split (\ref{di}).
There is also the error due to the Born-Oppenheimer approximation. It follows
that a more precise upper limit could be reached if the correction of the
minimal length may be included in the currently more developed quantum
molecular methods.

Nevertheless, the result (\ref{ub}) can at least be compared to those obtained
in the study of fundamental problems of quantum mechanics. The upper bound
(\ref{ub}) is weaker than the ones obtained in Refs.
\cite{brau,sandor,stetsko} for the hydrogen atom. There, the minimal length
has been supposed to be related to the finite size of the electron; and an
upper bound of about $10^{-7}$-$10^{-6}$ \AA \ has been resulted by using the
high-precision experimental data for the transition 1S-2S and for the Lamb
shift. It follows that in the study of the vibrations of the molecule H$_{2}$
(motion of the nuclei), this elementary length would rather be a scale related
to the size of the proton not to that of the electron. It is therefore not
surprising to find such value to $\left(  \Delta X\right)  _{\min}$. In Ref.
\cite{chang}, the deformed spectrum of the harmonic oscillator has been
applied to the motion of electrons in a Penning trap, which led to an upper
bound of the same order of magnitude as in the hydrogen atom problem. However,
in Ref. \cite{brau2}, the investigation of the effect of the minimal length on
the energy levels of ultracold neutrons in the Earth's gravitational field,
led to an upper bound of about $24$ \AA , which is larger than the value
(\ref{ub}).

It turns out that in the study of quantum mechanical problems, this
fundamental length would not be close to the Planck scale, but it is rather a
scale related to the system being studied and would then depend on its characteristics.

\section{Summary and Conclusion}

We have studied the Kratzer's molecular potential $V(r)=g_{1}/r^{2}-g_{2}/r$
in quantum mechanics with a deformed Heisenberg algebra characterized by the
existence of a minimal length, $\left(  \Delta X\right)  _{\min}=\hbar
\sqrt{3\beta+\beta^{^{\prime}}}$. In the momentum representation, the
Schr\"{o}dinger equation was established in the particular case $\beta
^{\prime}=2\beta$ in the first order of the deformation parameter $\beta$. It
is then transformed to a canonical form of Fuchsian differential equations,
namely, a generalized Heun's equation, which reduces to a hypergeometric
equation for $\beta=0$ and to a Heun's equation for $g_{2}=0$. The behaviors
of the solutions show that the potential becomes regular even for values of
the coupling constant $g_{1}$ such as $2\mu g_{1}/\hbar^{2}\leq$ $-1/4$. In
coordinate space, an analytical expression of the energy spectrum has been
perturbatively derived in the first order of $\beta$. Through this spectrum,
we have explicitly shown the effect of the minimal length on the rovibrational
energy levels of diatomic molecules with KP interaction. The leading
correction introduced by the minimal length affects the energy of the
anharmonicity of vibrations, and the next-leading contribution concerns the
vibration-rotation coupling. However, the energy of harmonic vibrations, which
is the principal part of the vibration-rotation spectrum, is not significantly
affected, and the rotational levels is absolutely not corrected by this
deformed algebra. Furthermore, it has been outlined that the expression of the
deformed energy spectrum of KP can be viewed as a formula of a three-parameter
potential, and thus the parameter $\beta$ can be determined together with
those of Kratzer by fitting the spectroscopic data of diatomic molecules. We
have estimated an upper bound of the minimal length of about $0.01$ \AA \ by
comparing the theoretical and experimental values of the vibrational
ground-state energy of the molecule H$_{2}$. This result shows that the
minimal length might be in fact viewed as an intrinsic scale characterizing
the system under study, and not a parameter that accounts for quantum
gravitational effects, which are extremely small in molecular physics.

\begin{acknowledgments}
This work was supported by the Algerian Ministry of Higher Education and
Scientific Research, under the PNR Project No. 8/u18/4327 and the CNEPRU
Projects No. D017201600026 and D01720140007. I am grateful to my colleague A.
Boukhellout for his collaboration. I thank J-R. Cudell for his kind
hospitality during my visit to the "Service IFPA" at ULg, where part of this
work has been completed, and T. Birkandan for stimulating discussions on the
generalized Heun equations and for the warm hospitality while I was in visit
to the Istanbul Technical University. I apreciate my fruitful contact with P.
Hajigeorgiou. I acknowledge the referees for their valuable suggestion which
improved this manuscript.
\end{acknowledgments}

\end{document}